\documentclass[sigconf, nonacm]{acmart}

\usepackage{array}
\usepackage{multirow}
\usepackage{booktabs}
\usepackage{longtable}
\usepackage{multicol}

\usepackage{microtype}   
\usepackage{xurl}        
\usepackage{wrapfig}
\usepackage{fontspec} 
\usepackage{polyglossia} 

\usepackage{tabularx}
\usepackage{longtable}

\setdefaultlanguage{english}

\setotherlanguage{bengali}
\setotherlanguage{chakma} 

\AtBeginDocument{%
  }

\def\authnotes{1}
\newcounter{notectr}[section]
\newcommand{\thenote}{\thesubsection.\arabic{notectr}\refstepcounter{notectr}}
\newcommand{\newedits}[1]{\textcolor{black}{#1}}

\newcommand{\note}[2]{$\ll$#1~\thenote: #2$\gg$}
\newcommand{\cnote}[1]{\ifnum\authnotes=1 \textcolor{blue}{\note{Comment:}{#1}}\fi}

\setcopyright{none}

\usepackage{xcolor}
\usepackage{todonotes}

\usepackage{array}
\newcolumntype{F}{>{\arraybackslash}m{10cm}}
\newcolumntype{E}{>{\arraybackslash}m{12.8cm}}
\newcolumntype{D}{>{\arraybackslash}m{5cm}}
\newcolumntype{C}{>{\arraybackslash}m{3cm}}
\newcolumntype{B}{>{\arraybackslash}m{1.7cm}}
\newcolumntype{A}{>{\arraybackslash}m{1.5cm}}

\usepackage[normalem]{ulem}

\begin{document}

\title[Chakma Language Revitalization]{Re-educating Educated Ones: A Case Study on Chakma Language Revitalization in Chittagong Hill Tracts}


\author{Avijoy Chakma}
\affiliation{
  \institution{Bowie State University}
  \city{Bowie}
  \state{Maryland}
  \country{USA}
}
\email{achakma@bowiestate.edu}

\author{Adity Khisa}
\affiliation{
  \institution{University of Dhaka}
  \city{Dhaka}
  \state{Dhaka}
  \country{Bangladesh}
}
\email{bsse1334@iit.du.ac.bd}

\author{Soham Khisa}
\affiliation{
  \institution{University of Maryland Baltimore County}
  \city{Baltimore}
  \state{Maryland}
  \country{USA}
}
\email{skhisa1@umbc.edu}

\author{Jannatun Noor}
\affiliation{
  \institution{United International University}
  \city{Dhaka}
  \state{Dhaka}
  \country{Bangladesh}
}
\email{jannatun@cse.uiu.ac.bd}

\author{Sharifa Sultana}
\affiliation{
  \institution{University of Illinois Urbana-Champaign}
  \city{Champaign}
  \state{Illinois}
  \country{USA}
}
\email{sharifas@illinois.edu}


\renewcommand{\shortauthors}{Chakma et al.}

\begin{abstract}
Indigenous languages face significant cultural oppression from official state languages, particularly in the Global South. We investigate the Bangladeshi Chakma language revitalization movement, a community grappling with language liquidity and amalgamation into the dominant Bengali language. Our six-month-long qualitative study involving interviews and focus group discussions with Chakma language learning stakeholders uncovered existing community socio-economic challenges and resilience strategies. We noted the need for culturally grounded digital tools and resources. We propose an ICT-mediated community-centric framework for Indigenous language revitalization in the Global South, emphasizing the integration of historical identity elements, stakeholder-defined requirements, and effective digital engagement strategies to empower communities in preserving their linguistic and cultural heritage.

\end{abstract}


\maketitle

\section{Introduction}
Each language reflects a unique perspective on cultural diversity, capturing the experiences, traditions, and knowledge of its speakers \cite{andriolo2019international}. A mother tongue is fundamental, deeply rooted in personal and cultural identity, allowing for clear, accurate, and emotional expression,~\newedits{which is} crucial for children's cognitive growth \cite{nishanthi2020understanding}. Education in a familiar language enhances understanding, critical thinking, and academic performance \cite{orenhancing}. Of the world's approximately 7,000 languages \cite{ethnologue2023, crystal2010cambridge}, few receive official recognition; the rest face discrimination and endangerment. Non-official languages encounter systemic, cultural, and technological hurdles affecting their survival and social status. These challenges are especially acute for Indigenous and minority languages in the Global South, where national and global forces often marginalize non-dominant languages. While multilingualism is common globally, it often means a majority language alongside marginalized minority tongues ~\cite{lindell2025role}, \newedits{where minority tongues are often endangered by colonization, suppression, or assimilation.}

\newedits{Indigenous language research identifies language liquidity~\cite{androutsopoulos2015networked, blommaert2010sociolinguistics}, language amalgamation~\cite{bakker1997language, muysken2000bilingual}, and code-mixing~\cite{myers1997duelling} as some of the core phenomena that endanger minority languages. \textbf{Language liquidity} refers to fluid, adaptive practices that move across scripts, phonetic spellings, and emojis from multiple languages, and this fluidity gradually weakens the boundaries and stability of the minority language~\cite{androutsopoulos2015networked, blommaert2010sociolinguistics}. \textbf{Language amalgamation} describes the long-term merging of grammatical, lexical, and phonological features across languages, and this process reduces the distinctiveness of Indigenous linguistic systems~\cite{bakker1997language, muysken2000bilingual}. \textbf{Code-mixing} occurs when speakers insert elements from one language into another within a single utterance, and this insertion pattern accelerates the displacement of minority vocabulary~\cite{myers1997duelling}. Within sociotechnical systems, these dynamics intensify as social media and generative AI reward dominant-language use, pushing Indigenous creators, children, and youth toward linguistic shifts that diminish the everyday presence and status of minority languages \cite{tan2024function, siebenhutterminority, valijarvi2023role, rofi2025good}. These phenomena are particularly threatening for unprotected indigenous languages, particularly in the Global South. Note that a threat to one's language not only supersedes the community's oral and written communication but also endangers their history, myth, and morals, which are endangered by the dominant culture and the technologies used in their communications. Therefore, it is an HCI design responsibility to remain cautious of how platform architectures, AI systems, and everyday digital practices actively risk generating biased, non-inclusive data systems and AI tools, and/or participate in the erosion or preservation of Indigenous languages, making it necessary to investigate how communities experience, negotiate, and resist these pressures.}

\newedits{HCI has historically played a critical role in shaping how communities engage with digital tools for linguistic and cultural preservation, with prior work proposing strategies such as documenting spoken texts, grammar, and vocabulary; creating digital archives~\cite{moradi2020designing}; supporting Indigenous knowledge sharing~\cite{kotut2022winds}; and integrating storytelling and design guidelines~\cite{kotut2022winds, soro2015noticeboard, shiri2022indigenous}. Yet, it remains unclear whether these approaches foster sustained, everyday use among native speakers, especially in the Global South. For the Chakma community, Bangladesh's largest Indigenous group, yet only 0.003\% of the national population, the only existing minimal and community-led revitalization efforts are hindered by socio-political and economic barriers and limited institutional support. As HCI researchers, we aim to identify revitalization measures both aligned and mismatched with existing HCI literature while responding to community-defined priorities. We investigate the following research questions:}

\begin{quote}
RQ1: What systematic challenges does the Chakma language learning ecosystem face? \\
RQ2: What forms of resilience does the Chakma community exhibit against language amalgamation?\\
\newedits{RQ3: What revitalization measures do the Indigenous community members need to consider to complement the existing efforts in alignment with the existing HCI literature?}
\end{quote}

We conducted a six-month-long ethnographic study with the Chakma language learning ecosystem's stakeholders (e.g., instructors, students, and journalists, n=25) to understand current challenges, community resilience, the impact of language fluidity on amalgamation, and community perspectives on solutions. We found that the community experiences extreme human and material resource constraints and barely finds adequate support from the government and other non-government organizations. We noted that their resilience strategies involved finding ways for the Chakma language to coexist with the Bengali language, with influence and impact on the community's day-to-day language use and activities. Therefore, their designed Chakma language learning materials included commonly used Bengali definitions alongside Chakma definitions and explanations. Additionally, we found the tension of using different ICTs in the resource-constrained Chakma language learning ecosystem, while still promising to play supporting roles through documentation of lessons, fostering active engagement among the trainers and learners, and in the practice and assessment of their language learning. \newedits{All of these lead us to embrace the perspective of identity preservation through language revitalization and develop a language revitalization framework (MeILR) that addresses language-related challenges and operates as a countermeasure to decolonizing indigenous language.}

\newedits{We list our four key contributions as follows: \textbf{C1:} we provide a thematic analysis of rigorous interviews and focus group discussions with key stakeholders: language instructors, volunteers, students, social workers, religious workers, and journalists. \textbf{C2:} we identify the current preservation measures and practices, leading to insights into stakeholder needs and design guidelines for a prototype that empowers these communities in training, promoting, practicing, and eventually contributing to the language revitalization process through effective stakeholder engagement. \textbf{C3:} develop the methodological framework of indigenous language revitalization (MeILR) with a focus on an ICT-mediated, community-centric language revitalization process for Indigenous communities in the Global South, while including: (a) identifying existing historical documents that can enable speakers to draw upon moral, ethical, and societal values, considered identity elements, (b) determining community stakeholders' specific requirements, and (c) identifying effective digital mediums that would meaningfully engage stakeholders. \textbf{C4:} we discuss how MeILR-informed accessible digital tools can be created that address the specific needs identified by the community itself, thereby can be employed not only for the Chakma language preservation but also for other marginalized languages that aim to pursue a similar language preservation task. Together they significantly contribute to HCI subdomains such as Decolonial HCI (C1, C2, C3, C4), Human–Computer Interaction for Development (HCI4D) (C1, C3, C4), Preservation and Cultural Heritage HCI (C2, C3, C4), and CS-Ed / Learning Technologies (C1, C2, C3, C4).}

\textbf{Reflexivity Statement.}
\newedits{The first and the other two authors and ethnographers of this study are Chakma community members. They were born and raised in the Chittagong Hill Tracts, and are deeply familiar with the sociolinguistic histories, cultural practices, and political realities surrounding Chakma language revitalization. Our research and the data analysis are shaped by them. Their lived experience enabled trust-building, contextual interpretation, and culturally grounded translation, while also requiring continual reflection on how insider positionality shapes access, assumptions, and representation. Non-Chakma co-authors contributed by questioning taken-for-granted norms, supporting analytic rigor, and helping surface implicit meanings. Together, we approached this work as a collaborative, iterative process attentive to power, representation, and responsibility.}

\section{Related Works}

\textbf{Language revitalization} is both a linguistic concern and an HCI4D challenge, involving the design of systems that sustain identity and cultural continuity~\cite{hermes2012designing}. Research shows that institutional and community initiatives succeed when creating structured practice opportunities. Schools can expand experiential activities~\cite{nguyen2023preserving}, while clubs with games, debates, and storytelling foster engagement~\cite{educsci12110774}. Competitions further motivate minority students~\cite{Cham2020, Liu02012023}. Our review examines efforts across the Global North and South, situating \textit{Chakma} literacy as a design problem for technology-mediated interventions. \newedits{Note that throughout this paper we use the term “infrastructure” to refer to a combination and collaboration of digital tools (such as computers, laptops, projectors, mobile phones), shared community facilities (such as buildings, tables, and wi-fi), and the human labor and social arrangements required to operate, maintain, and support these resources.}

\subsection{Indigenous Language Literacy across the World}

In North America, a variety of approaches support Indigenous and heritage language preservation across institutional and community settings. In New Mexico, students use only the target language during class, supported by teachers who emphasize practice through singing, role-playing, paired conversations, translation activities, and vocabulary word walls that label classroom objects~\cite{khawaja2021consequences}. In Canada, the \textit{Miqqut} project promotes \textit{Inuktitut} literacy by embedding language learning in non-formal, culturally grounded activities such as traditional sewing, complemented by silent reading, vocabulary lists, and presentations guided by community elders—strengthening both language skills and intergenerational ties~\cite{tulloch2017transformational,tulloch2013miqqut}. Formal programs also contribute: the Canadian Indigenous Languages and Literacy Development Institute offers summer courses with Community Linguistic Certificates, and universities such as Alberta, Fraser Valley, and UBC provide degree-level Indigenous language instruction~\cite{khawaja2021consequences}. Beyond institutional efforts, early exposure to ethnic roots within families has been motivating for Korean-American youth, who often progress to volunteer work to further develop their heritage language~\cite{li1999can, park2007parents}. Positive parental attitudes and supportive interactions are shown to strengthen children’s identity, heritage language maintenance, and second-language development~\cite{wei1994three}. Korean immigrant parents in Montreal follow this practice, believing it helps children cherish their cultural identity~\cite{park2007parents}, while Korean media further supports language and cultural learning~\cite{park2023identity}. \newedits{These global efforts motivate our examination of the structural challenges within the Chakma learning ecosystem (RQ1) and how community-led practices may foster resilience in literacy development (RQ2).}

In South America, multiple efforts aim to improve Indigenous language literacy, yet discrimination, limited resources, and weak accountability remain widespread. In Mexico, the Nahuas (Nahuatl speakers) continue to face discrimination; children in Hueyapan are punished for using their native language and encounter barriers such as scarce educational materials and exclusion in job applications, pressures that push many toward Spanish or English despite government programs~\cite{olko2023spiral, hansen2010polysynthesis, olko2023spiral}. Peru has become a key site for revitalization, particularly in Amazonian regions. The Shipibo-Conibo communities in Peru and Brazil maintain their language through initiatives like the Shipibo Conibo Center, which leads youth radio programs and multimedia storytelling workshops~\cite{oyarce2019Indigenous}. Traditional songs and oral histories also sustain languages such as Asháninka and Aguaruna. Digital tools further complement these grassroots efforts—for example, QICHWABASE supports Quechua by harmonizing and making linguistic resources accessible online~\cite{huaman2023qichwabase}. \newedits{These South American initiatives prompt us to explore how Chakma communities resist linguistic erosion through their own cultural and technological practices (RQ2) and what revitalization measures can strengthen such efforts (RQ3).}

Efforts in Indigenous language revitalization in Asia are relatively limited and largely driven by government initiatives. Many programs use learners’ native languages as a medium of instruction. In Thailand, schools incorporate local languages alongside Thai to support education~\cite{darr2022minoritytongues}. Nepal implements multilingual education at the pre-primary level, using contextualized textbooks and local resources that make learning more engaging for children~\cite{shintan2018multilingual}. Another approach comes from studies applying Systemic Functional Linguistics (SFL), which emphasizes the connection between context, language, and meaning. Through peer review and indirect teacher feedback—such as guiding questions rather than explicit correction—students are encouraged to develop writing skills beyond grammar-focused instruction~\cite{zhang2018diluting}. \newedits{These regional approaches highlight persistent gaps in localized literacy support, reinforcing the need to identify the specific obstacles Chakma learners face and suitable revitalization strategies for their context (RQ1, RQ3).}

\subsection{History and Diversity of Language in CHT, Bangladesh}

The Chittagong Hill Tracts (CHT) in southeastern Bangladesh—comprising Rangamati, Khagrachari, and Bandarban districts is the home of thirteen Indigenous communities (Bawm, Chak, Chakma, Khumi, Kheyang, Marma, Murung, Pankhua, Lushai, Tanchangya, Tripura, Assamese, and Gorkha \cite{Mey1996, owp2022battered, dhamai2014overview}) and each maintains distinct traditions and languages spanning Indo-European and Sino-Tibetan families. Ethnologue identifies 41 `living languages' in Bangladesh \cite{ethnologue2023, crystal2010cambridge}; Chakma, Marma, Garo, and Santali are four major, yet are vulnerable to cultural dominance and a lack of support despite their rich heritage \cite{ethnologue2023}.

\subsubsection{History of Chakma Language} 

\begin{figure}[!t]
  \begin{center} 
    \vspace{-12pt}
    \includegraphics[width=0.4\textwidth, clip, trim={-2em 0 -2em 0}]{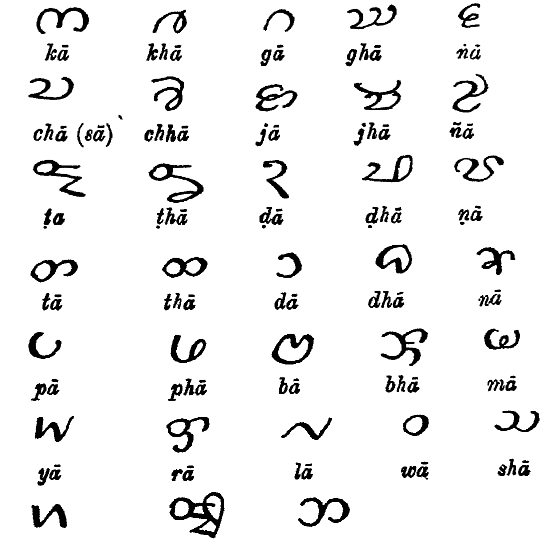}
  \end{center}
  \vspace{-1em}
  \caption{Chakma alphabet recorded in the~\textit{Linguistic Survey of India} \cite{Grierson1903}.}
  \vspace{-10pt}
  \label{fig:chakma_alphabets}
\end{figure}

Chakma people speak in Chakma language, and the Chakma language is the largest spoken Indigenous language in CHT. The \textit{Linguistic Survey of India} reported 33 Chakma letters \cite{Grierson1903, majeed2019colonialism} as depicted in Figure~\ref{fig:chakma_alphabets}. Despite phonetic similarities with the Bangla \cite{linguistic3paper} language, the Chakma script is cursive like Khmer and shows affinities with Arakani and Burmese scripts \cite{Grierson1903}. Historical artifacts suggest a strong past language literacy. In 1940, the manuscript \textit{Laldighir Pare Sagoto Badha Ghat} used Chakma scripts on palm leaves, which includes a Chakma folk-tale~\cite{Grierson1903}. Ancient religious texts such as \textit{Agortara}, which dates back to the 5th century AD, contain Chakma-scripted Theravada and Mahayana Buddhist elements \cite{Aghortara2006, linguistic12paper}. These indicate that writers, monks, teachers, and general people once practiced regular reading and writing. However, colonial and postcolonial politics and conflicts have reshaped this literacy trajectory. Despite historical evidence of literacy, today Chakma language education remains marginalized, demanding urgent revitalization. \newedits{This historical trajectory motivates our investigation into how present-day sociotechnical conditions shape Chakma language practices and community resilience (RQ1, RQ2).}

\subsubsection{Current Status of Chakma Language Literacy}

Reading, writing, speaking, and listening are the four essential skills for fluency, yet only a small portion of Chakma people in Bangladesh can read and write Chakma letters, and this number decreases further under time-constrained literacy tests. While most Chakma people can still speak and understand the language \cite{sumayra2023indigenous}, deep challenges persist in sustaining reading and writing practices. These challenges are historically and politically embedded. The British colonial period fractured the cultural and linguistic sovereignty of CHT, disrupted traditional governance and land rights, and accelerated cultural erosion \cite{chakma2024colonial, tofa2023trilingual}. Such systemic inequalities continued after the independence of Bangladesh, where institutional hegemony and nationalist policies marginalized Indigenous languages in education, media, and governance \cite{sultana2021ethnic}. Bangla was instituted as the state language, while English gained prominence as a tool for economic advancement \cite{tofa2023trilingual, afreen2020language, rahman2010multilingual}. This dual dominance has created significant obstacles for Indigenous languages, particularly Chakma.


Chakma children primarily use Chakma at home, but once enrolled in school, they must read, write, and learn exclusively in Bangla. Therefore, many children can speak Chakma but lack proficiency in its script \cite{sumayra2023indigenous}. Many young speakers also internalize stigma, feeling self-conscious about their Bangla accents, which further drives them to prioritize Bangla over Chakma \cite{afreen2020language}. The situation is exacerbated by the absence of curricular representation: textbooks rarely acknowledge Indigenous languages or cultural identities, eroding students' sense of belonging and harmony across communities \cite{ahmed2023linguistic}. When schools enforce instruction only in the dominant language, they enact what scholars call educational violence, denying Indigenous children their right to inclusive, high-quality education \cite{cueto2009explaining}. As a result, the dominance of Bangla and English systematically undermines Chakma literacy, weakens cultural sovereignty, and accelerates the broader crisis of language insolvency. \newedits{These contemporary challenges directly inform our inquiry into the structural barriers to Chakma literacy and the forms of resilience necessary for revitalization (RQ1, RQ2).} 

\subsection{Language, Historic, and Cultural Preservation in HCI}

\subsubsection{Indigenous Culture Preservation in HCI}
\newedits{HCI researchers have collaboratively designed socio-technical infrastructures and developed recommendations for preserving Indigenous knowledge, language, and culture~\cite{kotut2022winds, reitmaier2024cultivating,reitmaier2022opportunities, moradi2020designing, shiri2022indigenous}. Kotut and McCrickar examined Kalenjin community members' online and offline roles and proposed design principles that ensure elder participation, support younger members in translating and mediating traditional knowledge with the elders, retaining the contextual information of Indigenous knowledge to avoid misinterpretation or loss of meaning, and accommodate community consensus so that various versions of what are perceived as the original story can exist~\cite{kotut2022winds}. Other work focuses on creating accessible systems for low-resource languages. Reitmaier et al. co-designed an information retrieval system for a non-written language in India, addressing accessibility challenges in locating audio recordings by enabling photo tags and short annotated summaries~\cite{reitmaier2024cultivating}. Similar difficulties in retrieving voice messages were reported among Xhosa and Marathi users on WhatsApp~\cite{reitmaier2022opportunities}. Storytelling has also been a central design focus. Soro et al. co-created a community noticeboard with an Australian Aboriginal community to support oral and written storytelling, bilingual content~\cite{soro2015noticeboard}. Shiri et al. analyzed digital storytelling platforms in North America and Australia, highlighting the importance of oral forms of expression and the inclusion of audio and video as means of communication on these platforms~\cite{shiri2022indigenous}. Supporting decolonial aims, they emphasized the non-commercial, open-access design of the system to ensure that it remains under community control~\cite{shiri2022indigenous}. This work motivates us to identify revitalization measures that align with and extend HCI approaches to support Chakma language preservation (RQ3).}

\subsubsection{Culture, ICT and Decolonization in HCI}
\newedits{ICT-supported systems often overlook local cultural values and practices, inadvertently reproducing colonizing characteristics. Here, HCI scholarship emphasizes community-driven, culturally informed design and system development to ensure accessibility, trust, and resistance to colonial impositions~\cite{sultana2023communicating, sultana2019witchcraft, das2022collaborative}. Sultana et al.’s work in rural Bangladesh demonstrates how engaging with local moral worlds—such as beliefs around witchcraft—can foster communal relationships and support the co-creation of low-cost, situated health technologies~\cite{sultana2019witchcraft}. Another ethnography on everyday information visualization approaches and practices further shows that rural communities use culturally meaningful materials and symbols to convey risk, decision-making, and financial information—elements that Western visualization standards often fail to capture~\cite{sultana2023communicating}. Das et al. show that an online platform, BnQuora~\cite{bnquora2025}, facilitates “narrative resilience,” allowing users to challenge colonial narratives and reclaim agency through dialogic exchanges~\cite{das2022collaborative}. Dye et al.’s study of Cuba’s community-built StreetNet reveals how local computer networks can provide autonomy while requiring collective maintenance and negotiated access — valuable lessons for designing and developing resilient, community-oriented sociotechnical infrastructures~\cite{dye2019if}. These studies guide our examination of the sociotechnical obstacles Chakma communities encounter and the culturally grounded strategies they employ to maintain linguistic sovereignty (RQ1, RQ2).}

\subsubsection{Learning and Engagement in HCI}

\newedits{HCI research further explores how sociotechnical infrastructure can sustain engagement and provides guidelines for future development~\cite{Arawjo_2017, arawjo2019computing,Poon_2019}. Arawjo’s study of the Nairobi Play Project demonstrates how computing education can serve as a site for intercultural learning, where moments of friction, collaborative breakdowns, and unstructured interactions become productive spaces for engagement across cultural difference ~\cite{Arawjo_2017, arawjo2019computing}. Poon et al. designed and evaluated a quiz-based intervention aimed at helping school students in Cameroon and showed that students’ perceptions of personal security in the socio-technical environment and guardians’ attitudes toward technology shape engagement levels. They recommend clearer communication, adaptive quiz timing, culturally relevant content, and collaborative features to strengthen participation~\cite{Poon_2019}. They also recommend clearer communication with students and guardians, adaptive quiz timing, culturally relevant content, and collaborative features such as group quizzes, competitive feedback, and mechanisms to encourage students to discuss study materials with each other to strengthen participation~\cite{Poon_2019}. These findings inform our investigation into what revitalization measures can meaningfully support sustained engagement among Chakma learners within their sociotechnical environment (RQ3).}

\section{Methodology}


We conducted our study over six-month-long periods with a total of 25 participants. Our study comprises interviews, focus groups, and observations of three Chakma Language Learning Centers (CLLCs): Ridisudom, Gyanor Bareng, and Noyaram Chakma Sahitya Sangsad in the Rangamati and Khagrachari districts. Multiple authors, including the first author, belong to the local Chakma community as they were born and raised in this area. They also speak the Chakma language. Therefore, they have easy access to the community.

\begin{table*}[!ht]
\centering
\begin{tabular}{|c|c|c|c|c|c|}
\hline
Participant   & Age & Gender & Occupation & District & Role\\ \hline
P1 & 35-45 & Male & School Teacher & Rangamati & CLLC Educator\\ \hline
P2 & 45-55 & Male & Researcher & Rangamati & CLLC Supporter\\ \hline
P3 & 35-45 & Male & School Teacher & Rangamati & CLLC Educator\\ \hline
P4 & 35-45 & Male & Cultural Worker & Khagrachari & CLLC Educator\\ \hline
P5 & 65-75 & Male & Retired School Teacher & Rangamati & CLLC Educator\\ \hline
P6 & 35-45 & Male & Cultural Worker & Rangamati & CLLC Educator\\ \hline
P7 & 45-55 & Male & Social Worker & Rangamati & CLLC Supporter\\ \hline
P8 & 65-75 & Male & Poet, Writer & Rangamati & CLLC Supporter\\ \hline
P9 & 65-75 & Male & Poet, Writer & Rangamati & CLLC Supporter\\ \hline
P10 & 35-45 & Male & Journalist & Rangamati & CLLC Supporter\\ \hline
P11 & 18-20 & Male & Undergraduate Student & Rangamati & CLLC Trainee\\ \hline
P12 & 26-30 & Male & Undergraduate Student & Rangamati & CLLC Trainee\\ \hline
P13 & 18-20 & Female & Undergraduate Student & Rangamati & CLLC Trainee\\ \hline
P14 & 21-25 & Male & Undergraduate Student & Rangamati & CLLC Trainee\\ \hline
P15 & 21-25 & Male & Undergraduate Student & Rangamati & CLLC Trainee\\ \hline
P16 & 21-25 & Male & Undergraduate Student & Rangamati & CLLC Trainee\\ \hline
P17 & 21-25 & Male & Undergraduate Student & Rangamati & CLLC Trainee\\ \hline
P18 & 21-25 & Female & Undergraduate Student & Khagrachari & CLLC Trainee\\ \hline
P19 & 21-25 & Female & Undergraduate Student & Rangamati & CLLC Trainee\\ \hline
P20 & 21-25 & Male & Undergraduate Student & Rangamati & CLLC Trainee\\ \hline
P21 & 21-25 & Female & Undergraduate Student & Khagrachari & CLLC Trainee\\ \hline
P22 & 21-25 & Male & Undergraduate Student & Khagrachari & CLLC Trainee\\ \hline
P23 & 21-25 & Male & Undergraduate Student & Khagrachari & CLLC Trainee\\ \hline
P24 & 21-25 & Male & Buddhist Monk & Rangamati & CLLC Educator\\ \hline
P25 & 21-25 & Male & Buddhist Monk & Rangamati & CLLC Supporter\\ \hline
\end{tabular}
\caption{Participant's demographic information.}
\vspace{-20pt}
\label{tab:Demographic}
\end{table*}

\subsection{Participant Recruitment}

We recruited participants using snowball sampling, beginning with initial contacts made through three co-authors who are themselves members of the Chakma community. These co-authors reached out to local educators, cultural workers, and students who were already involved in CLLCs to explain the study's purpose and invite their participation. After each initial meeting, participants were asked to suggest others who might be interested, allowing us to expand the sample across educators, students, writers, poets, journalists, and community supporters. This method was particularly effective in building trust, as recruitment proceeded through existing social and cultural networks rather than external channels. We intentionally sought diversity across age, gender, occupation, and geographic background to ensure diverse perspectives. \newedits{We have conducted this study for six months, and three Chakma authors are local to this area. During participant recruitment, we found that the male representation among the educators is high, and we did not find any female educators.} All participants were informed about the voluntary nature of the study, their right to withdraw at any point, and the measures taken to ensure confidentiality and ethical participation. \newedits{Participants' demographic information is tabulated in Table~\ref{tab:Demographic}.}

\subsection{Interview}

We conducted a total of eleven semi-structured individual interviews with Chakma community members who were actively engaged in language learning, teaching, or cultural preservation. The participants included language instructors, cultural workers, retired schoolteachers, poets, writers, journalists, and undergraduate students across various disciplines. Five interviews were conducted online via Google Meet, with a co-author present during two of them to support interaction and manage technical issues, while six were conducted in person at locations convenient for the participants, such as their homes and local learning centers. Each interview lasted between 30 and 60 minutes. In these conversations, we asked about their experiences learning and teaching Chakma, the role of ICTs in language practice, and the challenges they encountered in sustaining literacy. Several participants also shared examples of teaching materials, personal writings, and social media posts to illustrate their responses. All interviews were conducted in the Chakma language, with informed consent obtained beforehand. We took detailed notes, and thirteen interviews were audio-recorded with permission for transcription and analysis.


\subsection{Focus Group Discussion}

In addition, we organized two focus group discussions to better understand collective perspectives on language revitalization. One session was held in person with seven participants, while the other took place online with six participants. The groups included instructors, students, and supporters of CLLCs, creating space for intergenerational and cross-role dialogue. Each focus group lasted about an hour and centered on the difficulties of sustaining Chakma language practice, strategies to overcome declining literacy, and ideas for using digital platforms to support learning. Participants actively exchanged views, compared experiences across educational and community contexts, and shared examples of both successful initiatives and ongoing struggles. \newedits{The sessions were conducted in the Chakma language.} Consent was obtained from all participants, and sessions were documented through notes and audio recordings, which were later transcribed and analyzed thematically.

\subsection{Data Collection and Analysis}
\newedits{The interviews and focus groups generated approximately 13 hours of audio recordings, which were securely stored in a restricted-access drive. Each file was assigned a unique identifier, later removed prior to analysis to ensure participant anonymity. The data corpus produced roughly 100 pages of transcription. The Chakma-speaking authors manually translated and transcribed the recordings. This manual process enabled careful interpretation of linguistic nuances, idioms, and culturally embedded meanings that automated tools might distort. Translations were repeatedly checked against the original Chakma audio for semantic fidelity. In instances where an exact English equivalent did not exist, we retained the Chakma term (see Table~\ref{tab:termsNewTable}) and provided an interpretive definition to preserve cultural integrity.}

\newedits{We employed thematic analysis \cite{boyatzis1998transforming} using an inductive coding (open coding) approach \cite{braun2006using}. Three team members independently coded each transcript, labeling data segments and identifying emerging categories. Initial coding began during early transcription to sustain an iterative link between data collection and interpretation \cite{glaser2017discovery}. Codes were compared across the team; when disagreements arose, they were resolved through collective discussion until conceptual consensus was achieved. Our research questions guided our analysis to ensure relevance to community-led education, local digital practices, and the everyday experiences of Chakma learners and facilitators. We did not rely on any predetermined theoretical lens or predefined themes, as our goal was to remain open to participant-driven insights. This openness allowed codes and categories to emerge organically from the data, reflecting the participants’ own interpretations and priorities. Initial codes included terms such as “Lack of institutional support”, “financial struggle”, “social hierarchy and awakening”, “challenges in phonetic learning”, “limited access to ICTs”, “lack of digitization and recording”, and “enforcing Chakma oral and written use.” After the coding stage, all team members collaboratively clustered related codes into broader themes that best describe our findings.}

\begin{table*}[!h]
    \begin{tabular}{|B|E|}
    \hline
    \hspace{1.8em}Terms & \hspace{15em} Definition\\ \hline
    \textit{Heng'gorong} & A bamboo-made musical instrument, resembling a jaw harp~\cite{Hengorong_2022, ChakmaJad_2017}.\\ \hline
    \textit{Ubogeet}
    
    & A traditional Chakma folk song, typically performed by a young man and woman to express their love and emotion, is distinguished by its distinctive tune, style, and expressiveness~\cite{MoanogharMusicSchool_2025, Chakma_2002}.\\ \hline
    \textit{Upojati} & Derived from Bengali, the term ‘sub-nation’ is used to label non-Bengali populations in Bangladesh, perceived with negative connotations by Indigenous peoples of Bangladesh who reject this designation, instead preferring Adivasi, meaning `Indigenous'. ~\cite{Partha_2024}. \\ \hline
    \textit{Bizu} & Bizu, the Chakma people’s most widely celebrated festival, is held April 13–15 to mark the end of the old year and the beginning of the new. \\ \hline

    \textit{Radhamon-Dhonpudi}  &  The greatest Chakma ballads - romantic stories of legendary heroic Radhamon and his beautiful lover, Dhonpudi~\cite{RadhamonDhanpudi_2004}.\\ \hline
    \textit{Gengkhuli}  &  Chakma ballad poet who used to perform ballads to entertain the village people in the past~\cite{RadhamonDhanpudi_2004}.\\ \hline
    \textit{Dagou-Hoda}   &  Words of wisdom - reflect elders’ guidance and cultural depth. The Chakma word \textit{Dagou}, derived from the Tibetan \textit{Dak} (‘knowledge’), also names a Buddhist community, underscoring linguistic and cultural ties between Buddhism and the Chakma people~\cite{Dewan_2005}.\\  \hline
    \end{tabular}   
    \caption{Definition of key terms.}
    \label{tab:termsNewTable}
\end{table*}

\subsection{Ethical Consideration}
The project is IRB-approved by the authors' university. We did not collect participants' personal information (name, email address, etc.) \cite{wiles2008management}. All participants voluntarily joined the meeting with their informed consent \cite{dixon2007beyond}. In the consent form and also at the beginning of the interview, one of the authors informed the participant that he/she could leave the meeting, skip, or refuse to answer any questions whenever necessary. Three Chakma authors, who were born and raised in Chakma villages and have spent more than 20 years in their native areas, were actively involved in questionnaire preparation, data collection, transcription, and analysis. All other authors contributed to reviewing and refining them.


\section{Chakma Language Preservation in CHT}
\subsection{Historic Context}
\label{sub_historic}

Based on our study, we found out that the Chakma Language preservation movement was unintentionally initiated by local Chakma \textit{Boiddo} (meaning physicians), and musicians. The participants highlighted how traditional healers served as custodians of the Chakma language and knowledge. Before the spread of modern medicine in the Hill Tracts, \textit{Boiddo} were central figures who treated a wide range of illnesses, from broken bones to cholera, and assisted in childbirth alongside \textit{oja-buri} (birth attendants). Importantly, their role extended beyond healthcare—they actively preserved the Chakma script by documenting treatments and remedies in written form. Most of their medical treatment and procedures are written in the Chakma alphabet in historical records. The participants explained how they helped with language preservation,



\begin{figure}[!t]
  \begin{center}
    \vspace{-20pt}
    \includegraphics[width=0.4\textwidth]{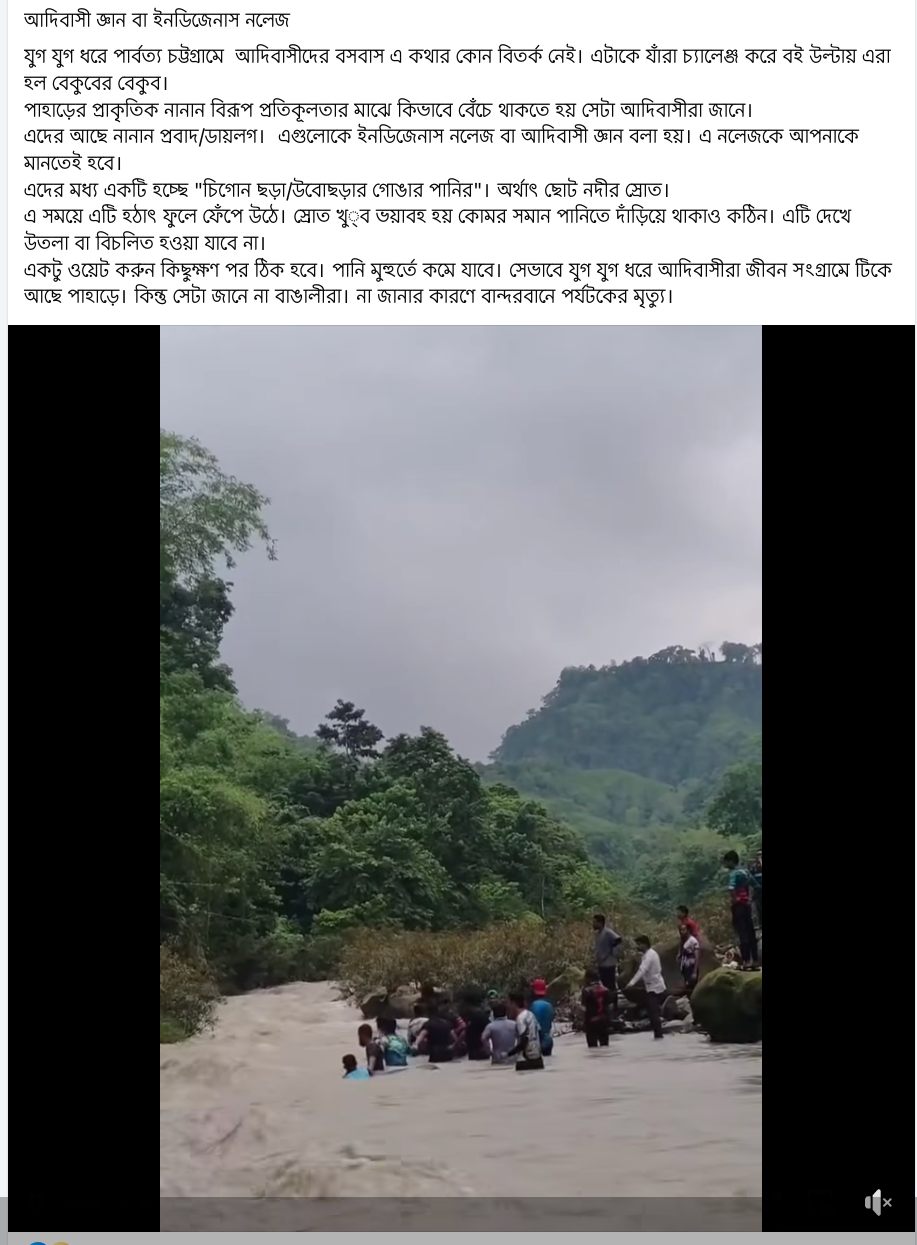}
  \end{center}
  \vspace{-10pt}
  \caption{Native Knowledge is preserved in the form of \textit{Dago-Hoda}, which is translated into idioms in English~\cite{FacebookReel2025}.}
  \vspace{-10pt}
  \label{fig:native_language}
\end{figure}

\begin{quotation}
\textit{``... another huge contribution made by the \textit{boddo} was in language preservation. They used to write in Chakma script, keeping written records in bags (\textit{jholas}). At first, they wrote on palm leaves (\textit{talpata}), and later moved to paper.''} [P8, Male]
\end{quotation}



Participants explained how cultural practices like \textit{Heng'gorong}~\cite{Hengorong_2022}
(a kind of Jaw Herp) and \textit{Ubogeet}~\cite{MoanogharMusicSchool_2025} (a kind of musical conversation between two persons) contribute to language preservation by embedding Chakma vocabulary, metaphors, and poetic forms into everyday expressions of love and social interaction. One of the senior participants mentioned,

\begin{quotation}
\textit{``Back in the day, young men and women expressed their romantic feelings through Hengorong. Another tradition is Ubogeet (folk songs), also centered on love and romance, where a man sings from one location and a woman responds from another (often across the hilly Jum cultivation fields). In this exchange, the woman sings a line, and the man replies, with both voices weaving imagery, metaphors, and poetic charm to express their affection.''} [P8, Male]
\end{quotation}




Additionally, Gengkhuli traditions are vital in keeping the Chakma language alive in ritual, song, and storytelling. Through long performances filled with unique vocabulary, metaphors, and poetic forms, they transmit linguistic richness across generations in ways that written texts alone cannot. Even as formal literacy declines, the performance of Gengkhuli songs ensures that younger audiences encounter rare words, idioms, and historical expressions, reinforcing continuity of language and culture in lived practice. The participants elaborated,

\begin{quotation}
\textit{``The Ubogeets are part of our historical tradition, tracing back through our wars and kings. We have a full history recorded, but unfortunately, this history is fading. Because we can become singers of Gengkhuli songs, but we cannot become Gengkhuli — that requires a level of
focus and also involves practice. Gengkhuli follows strict traditions; before singing, they set up the platform, worship it with wine and
money. The rituals invoke supernatural powers, and through these rites, the singer becomes ‘possessed’ and can perform the Gengkhuli songs continuously for seven days if desired...The ballads like Radhamon-Dhanpudi and the other historical ballads are also part of our cultural treasure, which is being lost.''} [P8, Male]
\end{quotation}



\textit{Dago-Hoda} (translates into Chakma idioms) is another element that preserves the native Indigenous knowledge. Some of these idioms are still relevant to this day and inform us about the nature of this region. Recently, several tourists died due to the flash flood while crossing a creek (a similar situation is depicted in Figure~\ref{fig:native_language}), which could have easily been avoided had the tourists known about some of the popular Chakma idioms. One was posted on social media,

\begin{quotation}
\textit{``Indigenous people have a proverb: ``Chigon chhara/ Ubo chhara’s gong’ar panir,'' which translates into ``the strong current of a small hill stream.'' At certain times, the stream from the hills suddenly swells up. The current becomes so fierce that even standing in waist-deep water becomes difficult. One must not panic or become restless seeing it. Just wait a little — the water level will drop in moments. In this way, generation after generation, the Indigenous people have survived. But Bengalis do not know this. Due to this ignorance, a tourist died in Bandarban.''} [P10, Male]
\end{quotation}


Together, these traditions demonstrate how Chakma language preservation is deeply tied to healing, art, and ecological survival. Even as formal literacy declines, such practices sustain linguistic diversity, transmit collective memory, and reinforce sovereignty through Indigenous knowledge systems.

\subsection{Current Indigenous Language Learning Infrastructure}

The Chakma community is one of the biggest Indigenous communities in the CHT area. The temples where Chakma-origin monks reside have not been historically active in teaching the Chakma language. Recently, there has been a growing tendency among the young monks to teach the Chakma language, primarily in the Rangamati district, but also observed in some other areas. Note that Chakma and Marma people follow Buddhism, and the primary religious textbook, \textbf{Tripitak}, is available only in Bangla, not in Chakma or Marma. Now, even the seasoned monks face difficulties in understanding and interpreting these Buddhist texts, due to the use of some obsolete Bangla words, and reported that \textit{Google Gemini} helped them understand better using much simpler terms and well-known common words. Some monks suggested translating the religious texts into the mother language because it would be easier to understand in their mother language, and not all Indigenous people can read Bangla. These monks inform that they felt a need to develop a workforce that can read and write Chakma, and thus, it motivates them to teach the Chakma language. We observe that temples can still play an important role. For example, one Buddhist monk has established a library to encourage young kids to read and motivate them to pursue higher studies. He has seen a positive response so far, whereas a similar role has been adopted by a layperson and failed miserably. The monk pointed out that it was due to the respect of the people towards the monk that led to the initiative's success.





\textit{Ridisudom}~\cite{Ridisudum} (one of the CLLCs) is in Rangamati. Currently, it operates from a rented floor in a multi-floor building and collects voluntary funds from the artists, teachers, and social workers to pay the monthly rent. Participants relied on a mix of traditional and digital resources to support language learning. Instructors primarily used printed books, handwritten notes, and board-marker setups in classrooms. However, learners do not have access to any tables during the paper-based evaluation process; instead, they use the floor-bed. Some temples also provide tables and study materials for their learners. On the other hand, a parliament member (MP) donated a computer to a CLLC called Noyaram Changma Sahitya Sangsad (NCSS)\cite{NCSS}, which uses this computer to teach the learners online. NCSS also has access to a printer and a projector. This organization also visits different schools and mostly uses a board-marker set up for Chakma language learning classes. Often, NCSS creates wall magazines and organizes study circles to encourage reading and writing practice among students.




Most organizations do not have access to Wi-Fi, and many learners who are still in elementary or high school do not have access to smartphones. We also observe that some learners studying at the college and from remote areas do not have access to smartphones. These learners use a traditional pen-paper-based approach to learn and practice the Chakma language. However, university-level learners have access to smartphones and desktop computers, and they use Facebook and WhatsApp Messenger to communicate with one another. They also use these social media platforms to practice the Chakma language through messaging and social media posts.

However, most CLLC instructors have personal smartphones and actively use social media platforms such as Facebook, YouTube, and WhatsApp. Some of these CLLC instructors maintain the social media handle for their corresponding organizations, such as Ridisudom, NCSS, and Gyanor Bareng. We have also observed university students using smartphones and desktop computers, whereas it is not common for undergraduate students studying in colleges. Most of the learners from the villages do not have access to smartphones or desktop computers. Organizations reported limited technological support, with many instructors and learners lacking smartphones or computers and relying on traditional chalkboard and marker teaching. For example, Ridisudom continues to use a board-marker setup in class. One participant highlighted that, despite having no smartphone himself, he personally knows individuals who lack access to basic digital tools, showing how limited access to these tools remains.  

\begin{quotation}
\textit{``Not everyone has a smartphone. Personally, I don't have a smartphone either. What I use is a desktop, which Basanti Di (Basanti Chakma, former MP) gave us for office use, and that's what I use. Maybe you'll be surprised to hear that in this modern age, it seems unbelievable! But the reality is, I personally don't have a smartphone. This is the situation.''} [P4, Male]
\end{quotation}


Additionally, there are other Indigenous language learning programs run by other Indigenous groups. For example, Marma is another ethnic Indigenous community in CHT. The Marma language has its own alphabet. Most of the Buddhist temples where Marma-origin monks reside play a dominant role in teaching the Marma language. Often, these temples maintain an orphanage, and the children are the primary learners of the temple-run language curriculum. Recently, branches of the Marma student organization, the Bangladesh Marma Student Council (BMSC)~\cite{bmsc}, at the University of Dhaka and the University of Chittagong, led a language learning campaign. One of the campaigns was telecast by a news channel called Jamuna TV, and it influenced the educated Marma students to learn their mother language.

\section{Findings}
\label{sec:findings}

Instructors at CLLCs focus on foundational literacy, beginning with the alphabet, pronunciation, and basic vocabulary. Lessons then extend to grammar, sentence construction, and reading comprehension. Students are also introduced to cultural forms such as poetry, folk songs, and proverbs, which help embed language learning within local traditions and identity preservation. While all of these are core to Chakma culture, the educators emphasized many different identity components in developing their teaching materials, challenges, and strategies to mitigate them.

\subsection{Challenges in Running CLLCs}

\subsubsection{Lack of institutional support}  

The CLLC instructors indicated a lack of interest in learning the Chakma language from the school-going students. The participants pointed out that, given there is no implicit requirement for knowing the Indigenous language and culture, as they are not part of the curriculum of employment exams for government service and other public and private sectors, learners are often demoralized and reluctant to go the extra mile. According to the CHT Accord (B.33.b, B.14)~\cite{PCJSS_CHTAccord1997} in 1997, the District Council (a government institution) is responsible for ensuring elementary education in the mother tongue and can create positions and appoint employees~\cite{guhathakurta2012chittagong}. However, no effective initiatives have been taken by the District Council that could ensure and improve Chakma language literacy among the Chakma people. CLLC instructors proposed that the Indigenous language- and culture-related questions should be part of government service and other public and private sector employment exams; however, this has rarely been practiced in reality. The CLLC instructors mentioned that they met with the District Council chairman and demanded that their strategy be implemented to encourage students to learn the Chakma language. Even though there is no conflict with existing laws and institutional policy, the District Council has not yet implemented this strategy.

\subsubsection{Financial Struggle} We also found out the financial struggles of CLLC organizations and how these constraints limit their capacity to expand language preservation work. The participants were frustrated with the lack of institutional support. They noted that although district councils have the capacity to support their initiative by allocating funding, the education department lacks the initiative to monitor and support such programs. As a result, grassroots organizations must rely on passion, informal donations, and ad-hoc community support rather than systematic government funding.

\begin{quotation}
\textit{`` We are planning to target the high schools that are located in the predominantly Chakma areas (chagala), for example - ‘Sapchori High School', where the majority is Chakma, especially students from Class 6 to 10. We've already discussed this with the chairman and the union council. If they provide us with a small amount of funding, at least if they can provide the transportation costs, then we can teach the students for free. This is our plan to grow student engagement. But we need support (ejal). We do not have minimum money.''} [P1, Male]
\end{quotation}

One of the strategies is voluntary work and minimal fees. To address the financial challenges with instructors' salaries, Ridisudom and NCSS instructors serve voluntarily during the campaign. At Ridisudom, the instructor roles are played by the teachers at the local primary school and social and cultural workers, who see this as a second voluntary job. To cover the printing costs of instructional and certification materials, these organizations charge the learners a small amount of 200 BDT (equivalent to 1.8 USD) as a one-time CLLC registration and enrollment fee. Additionally, they collect donations from various sources, but these are not sufficient, as the participants mentioned, 

\begin{quotation}
\textit{`` We have members; we have 144 members from various upazilas. We ask them to inquire at different schools, asking, 'We want to conduct a course on the Chakma language here, do you give us permission?' For works like these, a huge amount of money is required. But we don't have that much money, so we carry them out with the passion and determination we have for this work.''} [P4, Male]
\end{quotation}
     
Additionally, the participants mentioned that they occasionally received support from Indigenous businesspersons. For example, a local resort owner, Hemol Dewan (pseudonym, the owner of Ranga Dip resort), pledged to donate BDT 50,000 to support Chakma language education in a community club. The idea was to start with supporting a small number of learners and initiate a ripple effect,


\begin{quotation}
\textit{``Hemol Dewan has recently promised to donate 50k to teach Chakma writing in a club. With that 50k, if two batches can be trained, at least 60 people (30 in each batch) can learn to read and write in Chakma, and these 60 people can further encourage the other 60 people; this was his (Hemol) purpose. With this purpose, he helped (ejal) our club.''} [P1, Male]
\end{quotation}

The participants believed this investment would help more people gain the ability to read and write in Chakma. Hemol's contribution was therefore not only financial support but also a strategic investment in expanding community engagement with Chakma literacy.

\subsubsection{Social Hierarchy and Awakening}
We have interesting findings regarding the social class hierarchy and how it impacts the mother language learning efforts, thus the language revitalization efforts in general. For example, for youth in remote villages, the highest attainable goal is often securing a local teaching job under the district council. Since Chakma language skills might help in such recruitment, they see value in learning it. In contrast, urban, wealthier families prioritize English because their children dream of studying abroad and pursuing international careers.

\begin{quotation}
\textit{``Those of us who come from villages—our ultimate dream is to become a teacher under the district council. That is the highest sky for us.  Village people think: `If we get a teaching job in a primary school, and if Chakma language skills are needed, then I should learn it.' That's why we learned it—with the hope of getting a job. But people living in cities, from modern families, learn English. Because they dream of going abroad—to America, China, Japan—so they aim for scholarships. But can a child from a poor family living in the remote areas of the hills dream like that? No!''} [P10, Male]
\end{quotation}


We also noted the class-based limits on imagination: Poorer students in hill villages cannot imagine opportunities beyond their immediate surroundings. Their ``dream ceiling" is much lower than that of urban students, restricting the perceived utility of Chakma learning to local employment. Many educators also observed a sense of negative self-perception among students who are from remote areas. Some question the value of learning Chakma writing, while others appear to internalize socioeconomic limitations as personal failure. One participant reflected on how this mindset operates among children in the community:

\begin{quotation}
\textit{``What I particularly feel is that what works in our children's minds is the thought, 'My parents are poor, they can't read or write, and my father doesn't have a job, so I also won't be able to do much." This kind of thinking goes on in their minds, subconsciously. That's why they don't want to be involved with others and prefer to stay separate. So, we think they will get inspiration from the biographies of successful people. For example, recently, a girl (soccer player) named Monika has emerged from a similar background~\cite{StarOnlineReport_2019}.''} [P4, Male]
\end{quotation}


Thus, many students from poor families internalize their socioeconomic struggles, believing their parents' poverty and lack of education predetermine their own failures. This creates withdrawal and disengagement from learning. Educators respond by sharing success stories to inspire students and show that achievements are possible despite hardship.

\subsubsection{Challenges with Phonetic Learning}

We noted that instructors often rely on Bengali letters to approximate Chakma sounds, which means students learn the script through Bengali phonetic reference points rather than through Chakma's own sound system. As a result, instruction prioritizes writing as in how to represent Chakma or Bengali words in the Chakma script, as one of the participants mentioned, 

\begin{quotation}
\textit{``Another issue is that we're teaching our language mostly by pairing Chakma sounds with Bengali letters—like ka, kha, ga, gha—we just say them like that, and keep going. And the students are learning this way until the end of the course. That's why, when I asked someone, they said, ``Sir, we don't do that (proper pronunciation teaching). What we do is teach how to write a Bengali word in Chakma script, or how to write a Chakma word in Chakma script. We don't teach how to phonetically pronounce these sounds properly. Right now, we just teach the basics—nothing more.''''} [P9, Male]
\end{quotation}

This might cause neglect of proper pronunciation. Students, therefore, may gain only a surface-level familiarity with the script, without developing accurate oral fluency or deeper linguistic competence. This practice reflects the broader influence of Bengali dominance, where Chakma is taught as an overlay to another language rather than as a self-contained system. 

However, using the Bengali script to write Chakma may lead to mispronunciations and misunderstandings, because the two languages' phonetic systems do not align. The participants emphasized that the teachers highly recommend not using phonetic writing, as one of them explained,

\begin{quotation}
\textit{``Now we express in Chakma using Bengali script, but it is hard to read in the city or village because writing Chakma with Bengali letters is difficult for many. For example, we say ``ohring" (meaning deer), but if written in Bengali script as ``oh-ring," it won't sound right. But if written properly in Chakma script, it can be captured correctly. Writing Chakma in the Bengali script causes a lot of misunderstanding.''} [P8, Male]
\end{quotation}


Participants discussed forms of cultural encryption rooted in their socio-political histories, which require knowledge of Indigenous contexts to interpret. For instance, the Chakma term \textit{Cidere-Budara} literally refers to a toddler’s face smeared with food or random scribbles on the body, yet it is also used to describe a Bangladeshi law enforcement agency whose uniform evokes that imagery. Through such playful metaphors, Indigenous communities adapt everyday expressions to veil political critique. By likening state power to a child’s messy scribbles, the Chakma encode dissent in humor, fostering solidarity while remaining opaque to settlers and officials. This practice illustrates how language operates as a political tool of survival and agency.

\subsubsection{Limited Access to ICTs} 
\label{sub_limited_ICT}
\newedits{Participants shared a range of ICT limitations, including limited tech support, lack of digital teaching resources and tools. Participants also shared some of the resiliency efforts that they exercise within these limitations. We discuss these limitations and resiliency efforts in detail in the following-}

\paragraph{Limited Technical Support}  
\label{sub_tech_support}
Organizations reported limited technological support, with many instructors and learners lacking smartphones or computers and relying on traditional board-marker teaching. For example, Ridisudom continues to use a board-marker setup in class. On the other hand, donated computers allow some organizations, such as NCSS, to experiment with online teaching. However, participants noted that learners from rural areas often face severe connectivity issues. Even basic 2G services are sometimes unavailable, and damaged or sabotaged cell towers further restrict online meetings or regular participation.  

\begin{quotation}
\textit{``Those who are connected with the Noyaram Changma Sahitya Sangsad are mostly not from the city or the Upazila Sadar. Many of us are from Fhuhang para. They don't get a proper internet connection here, which is why they can't use Facebook or WhatsApp that much. If they need to use these apps, they might have to go to another place where there is internet.''} [P4, Male]
\end{quotation}

\paragraph{Lack of Digitization and Recording.}  
\label{sub_digitization}
Instructors often use printed books and Zoom but lack digitized resources. No classes are recorded, meaning dropout students cannot recover missed lessons. Several noted failed attempts at producing audio-visual materials due to a limited workforce and technical expertise. Others suggested that demonstration videos or audio pronunciations could improve learning outcomes.  

\begin{quotation}
\textit{``We used the printed books, but we could not manage the audio-visual medium. Many days ago, we tried to develop an audio-visual system for this course, but due to the lack of workforce and technical inexperience, we could not develop it....Our language course is not digitized.''} [P2, Male]
\end{quotation}

\paragraph{Limited Digital Tool Usage for Teaching.}  
Online classes still rely on cumbersome methods, such as drawing the alphabet with a mouse in a basic editor to replicate board-chalk learning. This slows instruction and discourages wider adoption.  

\begin{quotation}
\textit{``If we focus on the flow of the alphabet, how each letter should be written. A video demonstration would be easier to show on a projector/computer through slides. If a voice like the one used in the metro rail line is provided to pronounce our alphabet, it would be very helpful for pronunciation.''} [P4, Male]
\end{quotation}

\paragraph{Early Attempts to Use AI/LLMs.}  
Interestingly, some young instructors have started experimenting with AI tools. They record sentences daily to teach ChatGPT-like models Chakma vocabulary, aiming to preserve contextual meanings and humor.  

\begin{quotation}
\textit{``We can't find Chakma in AI technology. We are now trying to teach AI how to understand Chakma speech. We write down sentences daily. Every day, our technology is writing a book for learning the Chakma language. Since one is writing and the other is too, day by day, this way AI will be able to learn the Chakma language.''} [P7, Male]
\end{quotation}

\paragraph{Device and Font Limitations.}  
Many devices do not support Chakma fonts, preventing use even when users can write the script. Participants also described how unreadable fonts discourage online engagement—friends cannot read posts, leading to awkwardness, lack of interaction, or even ridicule. This pushes many toward Bangla, which ensures wider communication and validation.  

\begin{quotation}
\textit{``Before, I did not know how to write using the script, which is why I didn't. Now I can write, but my smartphone doesn't support the Chakma keyboard. It (OS) is an older version.''}[P15, Male]
\end{quotation}

\paragraph{Online Community Practices.}  
Despite these barriers, online groups provide spaces for collaborative learning. Messenger groups created after workshops allow learners to correct each other’s spelling and pronunciation. These peer-support networks help sustain language practice even without formal digital tools.  

\begin{quotation}
\textit{``We created a Messenger group for those who participated. So, if we face any issues with spelling, we ask questions there. Since several people participated in the workshop, there was always someone in that group who knew the correct answer.''} [P20, Male]
\end{quotation}

\subsection{Mitigation Strategy}
In our investigation, we find that these Indigenous Chakma community members adopt multi-faceted initiatives in their language revitalization effort, which includes the balanced use of the alphabet in social media and pressing governmental institutions.

\subsubsection{Enforcing Chakma Language in Oral and Written Communication} 
While most Chakma people maintain oral fluency, proficiency in reading and writing remains limited. Daily interactions with the Bengali population further accelerate the adoption of Bangla words, with participants estimating that 70--80\% of words in daily Chakma speech are Bengali. In response, some organizations pursue strong enforcement measures as part of language revitalization, framing the language not merely as a communication tool but as a marker of identity.  

\begin{quotation}
\textit{``Our organization is almost one and a half years old, and we have organized 15--16 programs. In these programs, we did not use Bangla fonts in the banner; rather, we used all Chakma fonts! It is up to you if you can understand it or not! If the invited guests do not understand it, then we will translate it for them. Also, we have a condition that if someone attends our programs, then he/she has to converse in authentic (jhora jhora) Chakma language.''} [P1, Male]
\end{quotation}

Such efforts, however, often meet practical challenges. Trainers observed that many learners come from poorer, rural backgrounds and carry feelings of inferiority. To address this, organizations supplement language enforcement with motivational strategies. For instance, instructors encourage writing through wall magazines and study circles, though participation remains low as students gravitate toward TikTok, mobile games, or card playing. To counter disengagement, some programs highlight the biographies of successful community members, showing that education and language pride can lead to tangible achievement.  

In addition, broader political contexts reinforce the symbolic value of learning Chakma. Participants emphasized that in times of communal tension and discrimination, the ability to assert an Indigenous identity through language strengthens claims to dignity and recognition. Yet, the same sociopolitical pressures, combined with economic precarity, often limit students’ capacity to sustain long-term engagement in language courses.  

\begin{quotation}
\textit{``Knowing our own language is extremely important for identifying ourselves as Indigenous people. It is essential for preserving our own existence. Because we often become victims of communal attacks and are labeled as `Upojati.' In such situations, we can respond by saying that we have our own language and fulfill all the criteria---so why should we still be called `Upojati' (less than a nation)?''} [P21, Female]
\end{quotation}

\subsubsection{Suggested Solution Strategies} 
\label{sub_sol_strategy}
Participants shared a range of strategies to strengthen Chakma language revitalization. \newedits{The founded strategies can be categorized into the following categories: 1) community-led activities, 2) leveraging social media, 3) improving writing practices, and raising awareness. We explain them in detail in the following-} 


\paragraph{Community-Led Activities.}
One strong recommendation was organizing grassroots competitions and cultural programs to inspire children. These could be tied to festivals such as \textit{Bizu}, where even small prizes can motivate participation. Several participants emphasized that such events do not require large budgets, only sincere community effort. 

\begin{quotation}
\textit{``If we want to make this effort work at the grassroots level, we can try organizing competitions — different kinds of language competitions. For example, during the Biju festival, language competitions are held. What if we arranged such contests in villages? Even if there are no clubs, people can still organize simple programs on special days. It doesn't take millions of taka — just a sincere, focused intention.''} [P10, Male]
\end{quotation}

\paragraph{Leveraging Social Media.}
Social media was viewed as both a challenge and an opportunity. Participants observed that content creators often default to Bangla for broader reach and monetization, discouraging Chakma-language production. Others suggested that a balanced approach—mixing Chakma with Bangla or English—could attract viewers while preserving linguistic identity. Small everyday practices, like writing one’s profile name in Chakma script or forming Messenger groups in Chakma, were also seen as important steps.  

\begin{quotation}
\textit{``What I understand is that those of us who learned the Chakma writing write more or less on social media. But we are not social media influencers; no matter how much we write, our content doesn't get that much reach. Indigenous influencers mostly use Bangla as their medium since it earns them more. So, if we create content in Chakma, how many people are actually going to watch it?''} [P12, Male]
\end{quotation}

\paragraph{Improving Writing Practices.}
A major concern was the lack of standardized grammar and the overreliance on transliteration, which often leads to errors in spelling and pronunciation. Teachers themselves sometimes transmit incorrect sounds, reinforcing mistakes at an early stage. Participants suggested developing structured frameworks and collaborating with elders—the custodians of oral traditions—to refine grammar, pronunciation, and meaning.  

\begin{quotation}
\textit{``Right now, just teaching one word at a time isn't enough. Where's the structure here? There's no proper framework — no grammar! We need to learn from the elderly. We need to approach them and say, ``Okay, I'm giving you a word, could you write it down for me?'' Then they'll write it, pronounce it, and explain the meaning. This kind of effort is necessary — until we reach the true custodians of our language.''} [P9, Male]
\end{quotation}

\paragraph{Raising Awareness.}
\label{sub:awareness}
Educated youth were identified as key actors for initiating a cultural renaissance. Participants stressed that university students must first become aware of their own linguistic responsibilities before going to villages to raise awareness. In areas with limited internet access, such efforts could preempt the linguistic mixing that often accompanies digital influence. Parents and community leaders were also viewed as crucial allies in instilling the belief that language preservation is central to cultural survival.  

\begin{quotation}
\textit{``Those who are studying at universities need to first become aware of themselves. Because if they are not aware of themselves, how will they raise awareness among others when they go to work at the root level? This way, we need to create a cultural renaissance. By going to different villages, we can raise awareness among the community leaders and parents, telling them that in order to preserve a nation, the language is extremely important.''} [P17, Male]
\end{quotation}

\subsubsection{Functional Digital Tools} 
\label{sec:sub_functional_digital_tool}
Participants frequently emphasized the need for practical digital tools to support the learning and everyday use of the Chakma script. Since learners often face challenges in remembering the alphabet, spelling, and correctness, they imagined how mobile applications and translation technologies could sustain their practice in meaningful ways.  

\begin{quotation}
\textit{``I think a mobile app could be really helpful. I would want it to tell me whether what I wrote is correct or not. It should follow some standards, for instance, be accurate, like 80 or 90 percent at least. Not online, an offline dictionary would be even more helpful.''} [P14, Male]
\end{quotation}

This reflection illustrates how feedback-oriented apps or offline dictionaries could directly address the challenges of language practice, providing learners with confidence and motivation.  

\begin{quotation}
\textit{``If there were an option to translate Chakma script into Bangla or English, more people would be encouraged to write in Chakma. I mean, it's not like 100\% of people will become encouraged; at least more than 50\% would have become encouraged.''} [P12, Male]
\end{quotation}

Translation was viewed as a bridge between Chakma and dominant languages, encouraging expression and making communication more inclusive.  

\begin{quotation}
\textit{``Well, right now, a lot of us are learning Korean, Chinese, Japanese, and also English, since it's an international language. So if we could build a translation app on Facebook, that would really help. Or, let's say we wanted to talk to foreigners—if we could translate our language into theirs, and theirs into ours, it would make communication easier for them too.''} [P16, Male]
\end{quotation}

This aspiration reflects how youth situate Chakma in a global linguistic ecosystem, seeing technology as a way to connect across both local and international contexts.  

\begin{quotation}
\textit{``People have asked me, ``You work on the Chakma language and Script, so why is your Facebook name in Bangla or English?" And I tell them, ``I checked that Facebook fails to find my name through search (if I use Chakma font). Let's say a person who is not on my Facebook friend list wants to contact me, but they do not have my phone number. Now their only option is to find me on Facebook, but then he won't be able to find me.''} [P11, Male]
\end{quotation}

Even basic social media features, such as search, were reported as exclusionary when names were written in Chakma, revealing structural limitations in digital platforms that discourage users from displaying their identity in their mother tongue.

\subsubsection{Cultural Media Solutions}  \label{sub_5.2.4}
Beyond functional tools, participants also recognized the importance of culturally grounded media to revitalize interest in Chakma learning. They observed that younger generations often resist traditional book-based learning and are instead motivated by digital media.  

\begin{quotation}
\textit{``This is exactly what I'm talking about – technology is needed. The current generation reads less and is averse to books. But if we could use technology properly, like creating cartoons, films, short films, or even cartoons with Chakma rhymes, it could make a huge difference.''} [P4, Male]
\end{quotation}

This perspective highlights how cultural media such as animations, rhymes, and films could not only capture the attention of youth but also embed the Chakma language into everyday entertainment. Such approaches would transform language preservation into an engaging, accessible, and collective practice that aligns with contemporary digital consumption habits.

\section{Discussions}
This study reveals how efforts to preserve and practice the Chakma language are deeply entangled with questions of cultural survival, infrastructural constraints, and digital mediation. Our findings demonstrate that language is not only a communicative tool but also a symbol of sovereignty, identity, and political resistance. Below, we discuss how our research contributes to HCI in both theory and design fronts.

\subsection{Design Implications and Challenges}

Our findings point to several implications for designing systems that support Indigenous language revitalization in the CHT. First, infrastructural inequalities remain a central barrier: many instructors and learners lack smartphones, rely on donated desktops, or face persistent connectivity gaps. This suggests a need for low-bandwidth, offline-first tools similar to KoboToolbox or Jolla’s Sailfish OS, which are designed for marginal infrastructures \cite{lakshminarasimhappa2022web, tzvetanov2020first}. Likewise, ongoing issues of font incompatibility and keyboard support mirror challenges faced in other Indigenous contexts, where initiatives like Unicode keyboards in West Africa or FirstVoices in Canada have been critical for uptake \cite{bodomo2006unicode, chase2022networks}. Embedding such infrastructural adaptability into design is essential for ensuring continuity in language learning. \newedits{However, Dye et al. show that community-owned tools that provide technical sovereignty, minimizing unjust outside restrictions, also pose occasional maintenance challenges for community members ~\cite{dye2019if}. Therefore, having community members skilled in this additional area is also an important aspect for a sustainable tech-mediated language revitalization.}

Second, design must negotiate the politics of language as identity. Some organizations enforce “authentic” Chakma-only use in banners and programs, asserting sovereignty but risking exclusion of learners with partial proficiency. Similar tensions have been observed in Māori revitalization apps \cite{ka2013new}, which scaffold varying levels of literacy through graduated exposure. Systems for the CHT must thus strike a balance: preserving symbolic strength while offering inclusivity and progressive learning paths. Youth practices also highlight the importance of meeting learners within popular digital cultures. As with Māori and Hawaiian revitalization efforts that leverage YouTube cartoons, TikTok challenges, and translation overlays, Chakma tools could embed playful content into mainstream platforms to foster everyday practice. Finally, emergent community-led experimentation with LLMs demonstrates an appetite for AI-enabled translation. Comparable initiatives, such as Masakhane for African languages, show both potential and risks around dataset bias and ownership \cite{malisa2017masakhane}. Co-curation of datasets and transparent governance will therefore be central for equitable AI adoption in Indigenous contexts.
\begin{figure*}[!t]
\centering
\includegraphics[width=0.8\textwidth]{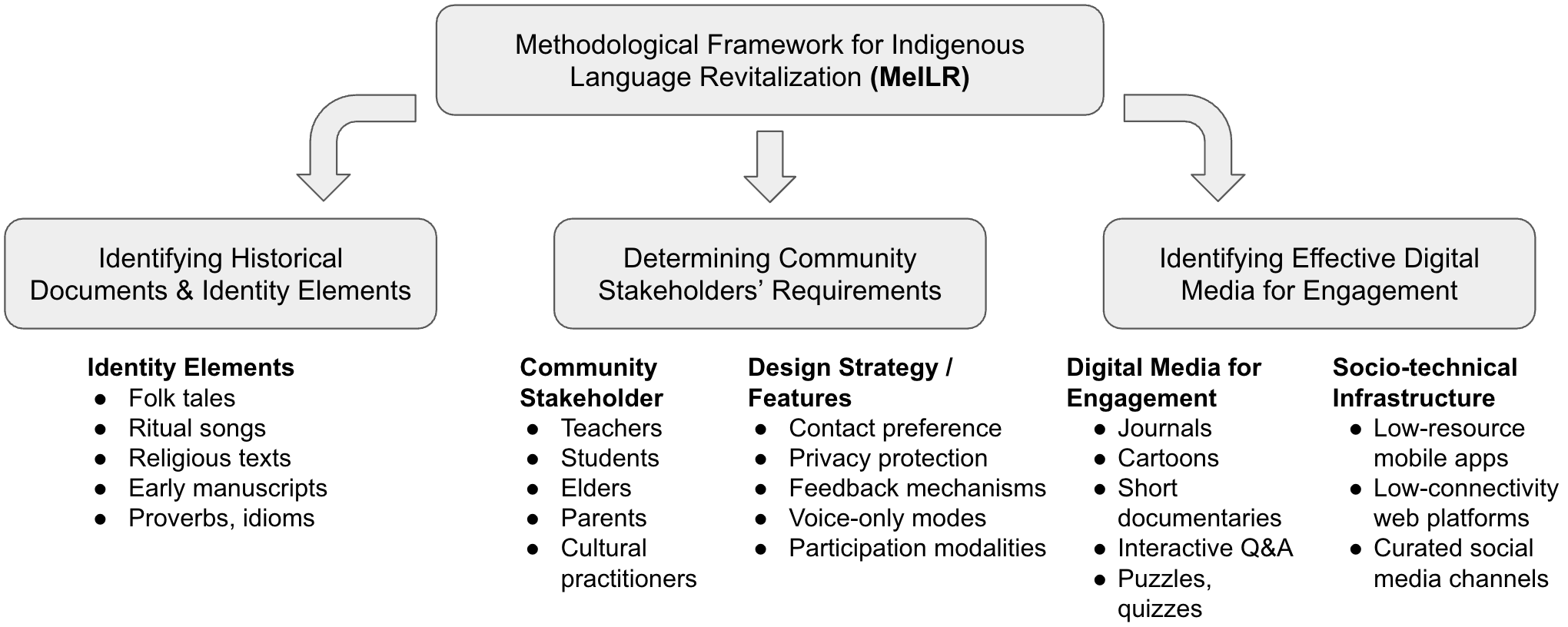}
  \vspace{-10pt}
  \caption{MeILR framework with its three pillars and corresponding parameters.}
  \vspace{-0pt}
  \label{fig:meilr}
\end{figure*}

\subsection{Language, Sovereignty, and Imagined Communities}

This work extends HCI4D scholarship's ongoing discourse of sovereignty by putting Benedict Anderson's idea of \textit{imagined communities} into conversation with Indigenous language revitalization in the CHT, foregrounding language as infrastructure for sovereignty \cite{anderson2020imagined}. Anderson shows that shared language—especially when stabilized through literacy and circulation—lets dispersed people imagine themselves as a collective. In our case, the erosion of Chakma script literacy and Bangla-dominant platforms constrain that imagination. \newedits{Especially, Chakma youths intentionally but unwillingly use Bengali in their online content to reach a wider audience, which would result in higher turnover from the content.} Revitalization efforts, therefore, treat Chakma not merely as a medium of exchange but as a sovereign sign: mandating ``jhora jhora" Chakma at events (see 5.2.1), insisting on Chakma-only banners even when translation is later required, and cultivating script use online. These are intentional boundary-making practices that assert political existence against linguistic assimilation.

\newedits{Our findings reveal that the erosion of Chakma script literacy does not simply limit participation in a shared reading public; it also reshapes how the community imagines itself across time and space. As learners rely on Bangla transliteration or fragmented digital supports, the imagined community becomes episodic rather than continuous, concentrated in pockets of connectivity and mediated by uneven device access, family attitudes, and local school constraints. These shifts alter the rhythms of collective imagination by breaking the slow, accumulative circulation that Anderson associates with shared print culture. Community-led efforts such as wall magazines, peer-support Messenger groups, and Chakma-only signage (see 5.2.1) therefore function as attempts to restore these disrupted temporalities by creating small, recurring sites of script circulation.}

At the same time, our findings deepen \newedits{existing HCI scholarship}, Sultana et al.'s account of digitally mediated belonging \cite{sultana2022imagined}. While they show ``imagined sisterhoods" in Facebook groups, we show how study circles, wall magazines, Messenger groups, and experimental AI/digital tools function as socio-technical infrastructures for an imagined Indigenous polity. These infrastructures counter isolation by letting learners see themselves as members of a sovereign linguistic community, yet they operate amid tensions—youth weigh Chakma literacy's returns against English/Bangla, rural connectivity lags, and platform logics reward non-Chakma content. Reading these frictions through HCI highlights design stakes: technologies that stabilize script, normalize bilingual visibility without subordination, and support low-bandwidth collaboration are not just pedagogical aids; they are instruments through which marginalized groups imagine, claim, and enact collective futures.

\subsection{Toward Methodological Framework for Indigenous Language Revitalization (MeILR)}

\newedits{Our findings show that revitalization in the Chakma community depends on raising collective awareness of linguistic erosion and creating spaces where learners can reflect on lived experience and challenge the socio-political forces that marginalize their language (\ref{sub:awareness}). This aligns with Freire’s Pedagogy of the Oppressed, which emphasizes dialogic, problem-posing learning that fosters critical consciousness and transformative action \cite{Freire_1970}. HCI scholarship similarly highlights the need for dialogic engagement to reclaim narrative agency in colonial contexts \cite{das2022collaborative} and warns that collaborative systems can reproduce structural exclusion \cite{fox2017social}. Together, these perspectives indicate that revitalization requires sociotechnical arrangements that prioritize dialogue, agency, and community-defined participation.}

\newedits{Building on this foundation, our ICT-mediated Methodological Framework for Indigenous Language Revitalization (MeILR) framework confronts the colonial and infrastructural conditions that suppress minority languages while cultivating the participatory dynamics necessary for linguistic resurgence. MeILR embeds Indigenous leadership, supports community ownership of knowledge and data, and treats language as a site of agency rather than deficit. The framework moves beyond documentation toward active, sustained engagement that addresses pressures such as language liquidity \cite{androutsopoulos2015networked, blommaert2010sociolinguistics} and language amalgamation \cite{bakker1997language, muysken2000bilingual}. MeILR is structured around three interconnected pillars: identifying historical identity elements, eliciting community stakeholders’ requirements, and selecting effective digital mediums for engagement.}

\subsubsection{Identifying Historical Documents and Identity Elements}

\newedits{The first pillar of MeILR emphasizes integrating historical documents and identity elements—oral histories, folk narratives, songs, rituals, and early records—as ``repositories of collective identity'' rather than mere linguistic artifacts. As Sultana et al. show, rural visual narratives in Bangladesh are polysemic, shaped by history and memory, and technology use in villages is entangled with moral and spiritual orders, underscoring the need for contextualized, culturally sensitive digitization~\cite{sultana2023communicating, sultana2019witchcraft}. For the Chakma community, this entails drawing on folk tales, religious texts, and customs linked to linguistic expression (\ref{sub_historic}), ensuring that revitalization is not ``decontextualized but culturally grounded''~\cite{sultana2019witchcraft}. Such rooting helps counter \textit{language amalgamation} by reinforcing unique cultural nuances~\cite{bhabha2012location, motsaathebe2018subaltern, kumaravadivelu2016decolonial}. From an ICT perspective, this requires careful digitization, annotation, and archiving for intergenerational transfer, while addressing challenges of fragile records, cultural protocols, and intellectual property rights.}
\begin{table*}[!t]
\centering
\begin{tabular}{>{\raggedright\arraybackslash}p{2.3cm}
                >{\raggedright\arraybackslash}p{6.5cm}
                >{\raggedright\arraybackslash}p{4.8cm}}
\hline
\textbf{Parameter} & \textbf{Value} & \textbf{Example Values and Justification} \\
\hline
Socio-technical Infrastructure 
& Low-resource mobile app, low-connectivity
web platforms, curated social media channel 
&  Allows to capture human-human component, social ties, technical component\\
\hline
Identity Elements 
&  Folk tales, ritual songs, religious texts, early manuscripts, poems, short stories, proverbs
& Serve as ``repositories of collective identity''\\
\hline
Digital Media for Engagement 
& Journal, cartoon, short documentaries, storytelling, discussion
session with interactive Q/A settings, asking quiz questions, puzzles
& Aligns with participants’ preferences. \\
\hline
Community Stakeholder
& Teachers, students, elders, parents, cultural practitioners
& Ensures multi-perspective, community-driven revitalization. \\
\hline
Design Strategy/Features
& understanding participant contact preference, privacy preservation, feedback mechanism, interactive voice-only engagement,
& Ensures continuous engagement\\
\hline
\end{tabular}
\caption{Parameters and example values for an ICT-Mediated, Community-Centric Framework.}
\vspace{-20pt}
\label{tab:param}
\end{table*}

\subsubsection{Determining Community Stakeholders' Specific Requirements}

\newedits{The second pillar of MeILR emphasizes systematically identifying the needs of stakeholders such as instructors, volunteers, students, social workers, journalists, and writers. This extends HCI work that calls for co-designing systems that align with community values and practices \cite{Cox_2021, Poon_2019, reitmaier2024cultivating}. Cox et al. argue through Indigenous Standpoint Theory that ICT interventions must be driven by community aspirations \cite{Cox_2021}. Poon et al. show that participation in ICT initiatives depends on trust, credibility, and community acceptance \cite{Poon_2019}. Our findings highlight five recurring requirements: (1) translation functionality in mobile and social media platforms, (2) keyboard support for apps (\ref{sec:sub_functional_digital_tool}), (3) audio-visual content for alphabet writing and pronunciation (\ref{sub_digitization}), (4) affordable devices for marginalized users (\ref{sub_limited_ICT}), and (5) integration of cultural media such as cartoons, films, and rhymes (\ref{sub_5.2.4}). Prioritizing these needs ensures that digital tools remain usable, culturally desirable, and capable of resisting passive consumption that accelerates language liquidity.}

\subsubsection{Identifying Effective Digital Media for Engagement}

\newedits{The final pillar of MeILR focuses on selecting digital media that meaningfully engage stakeholders by being accessible, culturally aligned, and supportive of active practice. Effectiveness here depends not on technological sophistication but on cultural trust, accessibility, and fit with existing socio-cultural dynamics. Prior HCI work shows that culturally grounded computing can foster dialogue and identity reflection \cite{Rick_2021}, that gamified semantic scaffolding sustains engagement by making abstract concepts learnable \cite{Arawjo_2017}, and that systems in low-resource contexts require trust and accessibility to succeed \cite{Poon_2019}. Our findings echo these insights: participants emphasized embedding Chakma cultural content such as cartoons, films, and rhymes (\ref{sub_5.2.4}), leveraging local social media influencers to raise awareness, and linking digital tools with community-led village activities (\ref{sub_sol_strategy}). Such integration builds trust, supports intergenerational learning, and sustains motivation over time. Practical options in Global South contexts include mobile apps, low-connectivity web platforms, and curated social media channels that shift from static documentation to interactive learning environments. These media must integrate historical identity elements and fulfill stakeholder requirements identified in the prior pillars so that revitalization remains socially embedded and technologically supported.}


\newedits{To understand how MeILR would turn into a realizable action component, we envision several parameters (Table~\ref{tab:param}) presented in Figure~\ref{fig:meilr} that would holistically shape its best functionality. Note that these parameters would be identified, challenged, and reevaluated based on the contextual implementation, maintenance, stakeholders, and engagement strategy, which makes MeILR value-sensitive, adaptive, and expandable to language revitalization of other similar contexts.}

\section{Limitations and Future Work}

This study focuses on understanding the Indigenous language revitalization challenges and the resiliency efforts, where we have considered the Chakma language as a case study. In addition, we propose an ICT-mediated framework as a representative for Indigenous languages in the Global South. We acknowledge a number of limitations, and most of these limitations are beyond the scope of this study. First, although this paper focuses on Bangladesh, Chakma communities also live in India and Myanmar, where challenges and resilience strategies may differ, and our documented challenges and the resiliency efforts may not be comprehensive. Second, the CHT is home to thirteen Indigenous communities, with the Chakma being the most populous and in a relatively advantageous position. Just as Bengali has historically functioned as the dominant language for the Chakma, the Chakma language itself may function as a dominant language in relation to other Indigenous groups in the region. Given differences in resources, sociopolitical contexts, and community structures, each of these groups may encounter distinct challenges and cultivate their own strategies of resilience. 

In the future, we aim to co-design an ICT-mediated framework as proposed in this study and identify the parameters and parameter values. Further, we plan to perform a participatory evaluation of the proposed framework to measure its effectiveness in the language revitalization movement. We envision that such design and evaluation would guide towards creation of an engaging digital experience for the Indigenous population of the Global South. Thus, further promoting the positive and inclusive digital media usage experience~\cite{rofi2025good}.


\section{Conclusion}

Our study shows how Chakma language revitalization efforts are shaped by infrastructural limits, digital exclusions, and political struggles over identity. By situating language as both communication and sovereignty, we highlight design pathways for inclusive, community-driven tools that sustain everyday practice. In doing so, we extend HCI’s role in supporting Indigenous futures through culturally grounded and sovereignty-aware technologies.


\begin{acks}
anonymized.
\end{acks}

\bibliographystyle{ACM-Reference-Format}
\bibliography{ref}

\end{document}